# Publishing an Article:
## A Main Goal for a Graduate Course

Michael Hilke
Department of Physics
McGill University
Montreal
Quebec, Canada
hilke@physics.mcgill.ca

*This essay discusses the effectiveness of a graduate course for physics students, with a course goal to write a publishable article on a modern research topic (graphene). I analyze the tools used to this end, which included Web 2.0 methods, in-class discussions and presentation, as well as extensive peer-review. In addition to producing a published article, this course led students to not only advance their mastery of a modern research topic but also to significantly improve their writing, discussion and presentation skills.*

# Introduction

In most graduate schools, be it at the masters or doctorate levels, students are required to pass several graduate courses or seminars, despite insidious comments by some professors that such courses are is simply a waste of time for the students, arguing that class-time takes time away from performing research. While this is debateable, it is nonetheless important to examine what the main objective of a graduate course is. When flipping through various existing graduate course objectives, two themes tend to stick out: (1) learning about current or recent research and (2) improving skills in academic writing, presentations and discussions. This drives the question of how can this be achieved most effectively? The solution that I present below discusses the requirement to include a publication as the outcome of a graduate course.

## Course Description

The following analysis is based on a one semester 2011 physics graduate course taught in 2011 (*Phys 634* at McGill) with an enrolment of 10 students. Most of the students were first year master's students. The topic was chosen because of tremendous current interest in the research community as illustrated by about 6000 articles published on graphene related research in 2011 alone (Source: *Web of Science*). This field of research exploded in 2004 thanks to its discovery that year by Geim and Novoselov, who were subsequently awarded the Nobel prize in 2010. Such a topic is commonly referred to as "hot" and is of interest to many students. While the sheer amount of publications in this field can be daunting for a professor, it is not difficult to imagine the effect on a first year graduate student. Nevertheless, I made the decision to offer a graduate course on this "hot" topic with the goal to publish a review article on this field, while restricting the focus on the experimental properties. The agreed upon title eventually became: *Experimental Review of Graphene* (Cooper et al., 2012).

I will argue below that such an end-goal for a graduate course, addresses the factors, which lead to an excellent learning outcome in terms of (1) learning about current or recent research and (2) writing and presentation skills). In addition, it addresses the "waste of time" argument in a powerful way, since it's straightforward for a graduate student to justify spending time on a course if it leads to a publication.

# Modern Science

## Research

Learning about current research is often done effectively starting at the undergraduate level, when students participate in research directly. This is justified, since involving undergraduate students in research is generally argued as leading to significant educational benefits (Gates, et al., 1999; Kardash, 2000; Zydney, et al., 2002; Seymour, et al., 2004). Research at the undergraduate level can take on several forms, such as an original contribution in a laboratory environment or an extensive literature review with a synthesis of the main results, which can be seen as a significant research form on its own (Cooper, 1988). At the master's level, the literature review is typically required for the thesis and often complimented by an original contribution. For a more conventional course setting, where a single instructor is in a classroom with many students, it is very difficult to conceive a course, where original research, like in a laboratory environment can be achieved for a large student enrollment. By contrast, a research level literature review can easily be implemented in a one instructor environment.

## Literature review

However, postgraduate students, do not necessarily understand the concepts involved in writing a literature review, as they often underestimate the need for critique in existing literature (Bitchener & Madeline 2007). As a way to introduce this research form, the starting point for our course *Phys 634*, was to have the students find all the existing reviews on the subject (in this case graphene). As a by-product, this enables an interesting discussion on various search tools and the use of up-stream and down-stream citations. While the students in class found more than 50 existing reviews, most of them were very focused in nature and it was quickly realized that there was none that was a comprehensive review of the experimental properties of graphene, which was the overextending topic of the course.

## In-class discussions and presentations

The next step was to define a table of content (*TOC*). The most valuable educational aspect, was the active participation induced by discussing the possible topics to be included in the *TOC*. This immediately provided a natural framework for active in-class discussions. The key is to come-up fairly quickly with a *TOC* so that the substantive work can start, but at the same time it is important to value the inputs of all students, since they will spend a considerable amount of time with the topics included in the *TOC*. Once the *TOC* is defined, the content constitutes the scientific material to be covered in the course. This was followed by the division in sections, where each section was put under the authority of one of the enrolled students. The first step for each student consisted in elaborating the content of the corresponding section and then to defend it in-class by means of a presentation. It is a little bit like a research defence, where the presenter has to introduce the different sub-topics in the literature review and defend their inclusion.

## Peer review

At this stage, the peer-review process really started, at first, through the critical discussions occurring during the presentation, and later through the peer-review of each section by two randomly chosen students in class. While peer review was shown to be an effective method to improve writing skills (Berg et al., 2006; Boud, Cohen, & Sampson, 1999; Falchikov, 1995; Stefani, 1994; Topping et al., 2000), in our context, peer review was also used as a method to learn and understand the material. In a way, the reviewer plays the role of the student learning new material, while providing critical feed-back, whereas the reviewed student gets the critical feed-back, which in turn, leads to improved writing and understanding outcomes (Lin, et al., 2001, Topping et al., 2000; Plutsky & Wilson, 2004, Xiao & Lucking, 2008).

## Web 2.0

To increase the effectiveness of the peer-review process, extensive use of *Web 2.0* was made. *Web 2.0* is the interactive form of the internet. Because of its interactive nature, in particular between students and also between instructors and learners, *Web 2.0* has gained considerable traction in online education (Beldarrain, 2006). *Web 2.0* includes many different tools, including course web sites, blogs, broadcasts, and various forms of web-sharing tools such as *wiki*. *Wiki* is studied extensively for its effectiveness in teaching subject matters that involve collaborative efforts such as the one found in mutual peer-review and collective knowledge construction (Bold 2006; Xiao & Lucking, 2008).

While most studies have focused on *wiki* designs, there are now many different tools available, which allow anyone to review and modify the content on a collection of documents on the web, both synchronously and asynchronously while enabling collaborative content-building for all participants. We used *Google* based web sharing software, including its web page creation tool as well as document sharing software (*Google docs* now replaced by *Google drive*), where each participant can add, comment and edit each document. Comments and changes were made directly on the shared document, where all changes can be traced by date and authorship, which also allows for an effective evaluation tool for the instructor.

The peer-review process, really works along two lines, the commenting tool, which is very effective in terms of the initial peer-review, whereas the editing tool is important for the later stages, which involves the creation of a unified document. For example, one student would use comments, to provide feed-back on the content and the writing style of a given section, while she would edit the tables of the entire document to make sure that they all follow the same formatting. Another student would be responsible for the formatting of the figures, while yet another for organizing the references. Many of these tasks are made simpler by using a *Latex* based word editor, which is comprised of one text file, containing all the written text, one reference file containing all the citations and another folder containing all the figures. The complete *pdf* output file (including all figures and citations) is obtained, using either a web - or home computer - based complier. Hence, not only does *Web 2.0* provide for an interactive framework for peer and instructor assessment but also for collaborative content building, involving a large number of collaborators, where the feedback is immediate and multidimensional.

A large portion of the peer-review process is happening electronically outside of the classroom, while following a shared timetable with imposed deadlines. At the end, all sections of the document are peer and instructor reviewed between five and ten times while some sections require large amounts of rewrites, which is done under the responsibility of one of the in-class peer-reviewers.

## Publishing

The last step involves getting the whole document into a publishable form. This is clearly an exciting moment for most of the students, since very few of them have published an article before. Moreover, the process contains many educational benefits. The first step is to decide which journal to publish in. Here the quality, the audience, and the ethics of the various journals and publishers can be considered. An important deciding factor for the students in class was to use journals, which are either open access or belong to non-profit organizations. The document was further geared towards beginning researchers in the field, which also restricts the number of possible journals. The decision was made to first attempt a submission to a top-tier journal, which could also lead to some high quality editorial feed-back, before eventually submitting to a second-tier journal in case the reviews turn out to be too critical.

Another interesting educational side-benefit related to the publishing process, is the discussion on copyright and plagiarism as well as defining each individual's contribution. It is helpful, that most publishers have well defined guidelines, which require students to familiarize themselves with these. Moreover, the students obtained feedback from other researchers in the field (including referees and editors), which they had to address in the resubmitted version.

## Conclusion

The produced document was subsequently accepted for publication in a peer review open access scientific journal (Cooper et al., 2012) and co-authored by all students and also posted on an electronic archive server (*ArXiv*). The evaluation of the students was based on the quality of the writing and the presentation, the diligence in the review process, and the overall participation. The learning outcome of the students was spectacular in terms of their confidence on the topic. This is largely because of the large amount of discussions which occurred in class and online on various topics of the document. The student's self-confidence got a further boost by the acceptance letter of the journal. The course evaluations by the students were excellent.

## References

Beldarrain, Y. (2006). Distance education trends: Integrating new technologies to foster student interaction and collaboration. *Distance Education, 27*, 139-153.

Berg, I. V. D., Admiraal, W., & Pilot, A. (2006). Peer assessment in university teaching: Evaluating seven course designs. *Assessment and Evaluation in Higher Education, 31*(1), 19−36.

Bitchener, J. & Banda, M. (2007). Postgraduate students' understanding of the functions of thesis sub-genres: the case of the literature review. *New Zealand Studies in Applied Linguistics, 13*(2), 89 –102.

Bold, M. (2006). Use of Wikis in Graduate Course Work. *Journal of Interactive Learning Research, 17*(1), 5-14. Chesapeake, VA: AACE.

Boud, D., Cohen, R., & Sampson, J. (1999). Peer learning and assessment. *Assessment and Evaluation in Higher Education, 24*(4), 413−426.

Cooper, D. R., D'Anjou, B., Ghattamaneni, N., Harack, B., Hilke, M., Horth, A., Majlis, N., Massicotte, M., Vandsburger, L., Whiteway, E., & Yu, V. (2012). Experimental Review of Graphene, *ISRN Condensed Matter Physics*, 501686, 1-56.

Cooper, H. (1988). Organizing Knowledge Syntheses: A Taxonomy of Literature Reviews. *Knowledge in Society* Spring: 104–126.

Falchikov, N. (1995). Peer feedback marking: Developing peer assessment. *Innovations in Education and Training International, 32*(2), 175−187.

Gates, A.Q., et al. (1999). Expanding participation in undergraduate research using the affinity group model. *Journal of Engineering Education, 88*(4), 409–414.

Kardash, C.M. (2000). Evaluation of an undergraduate research experience: Perceptions of undergraduate interns and their faculty mentors. *Journal of Educational Psychology, 92*, 191–201.

Lin, S. S. J., Liu, E. Z. F., & Yuan, S. M. (2001). Web-based peer assessment: Feedback for students with various thinking-styles. *Journal of Computer Assisted Learning, 17*, 420−432.

Plutsky, S., & Wilson, B. A. (2004). Comparison of the three methods for teaching and evaluating writing: A quasi-experimental study. *The Delta Pi Epsilon Journal, 46*(1), 50−61.

Seymour, E., A-B. Hunter, S.L. Laursen, and T. Deantoni. (2004). Establishing the benefits of research experiences for undergraduates in the sciences: First findings from a three-year study. *Science Education, 88*(4), 493–534.

Stefani, L. A. J. (1994). Peer, self and tutor assessment: Relative reliabilities. *Studies in Higher Education, 19*(1), 69−75.

Topping, K., Smith, F. F., Swanson, I., & Elliot, A. (2000). Formative peer assessment of academic writing between postgraduate students. *Assessment and Evaluation in Higher Education, 25*(2), 149−169.

Xiao, Y. and R. Lucking, R. (2008). The impact of two types of peer assessment on students' performance and satisfaction within a Wiki environment," *The Internet and Higher Education, 11*, 186-193.

Zydney, A. L., Bennett, J. S., Shahid, A., & Bauer, K. W. (2002). Impact of undergraduate research experience in engineering. *Journal of Engineering Education, 91*(2), 151–157.

# Biographies

Michael Hilke is Associate Professor of Physics at McGill University. He is the head of the Quantum Nano Electronics Laboratory, which involves many undergraduate and graduate students in his research on advanced materials and quantum engineering.